# SECONDARY EMISSION GAS CHAMBER

V. In'shakov, V. Kryshkin, V. Skvortsov

Institute for high energy physics, Protvino, Russian Federation

**Abstract**

For a hadron calorimeter active element there is considered a gaseous secondary emission detector (150 μm gap, 50 kV/cm). Such one-stage parallel plate chamber must be a radiation hard, fast and simple. A model of such detector has been produced, tested and some characteristics are presented.

## 1. Introduction

With the growth of collider luminosity and dimensions of detectors the requirements to radiation hardness, counting rate and cost of detectors come to the forefront. The micromesh chambers (see, for example, [1-3]) demonstrate the direction of the activity in solving the task. The example of utilizing such direction in calorimeters is given in [4-7]. On the other hand the potential capabilities of the secondary emission detectors attract attention for a long time [8-10]. We consider a detector that is some elaboration of a micromegas chamber - a secondary emission chamber with an amplification gap to increase counting rate and to simplify the construction.

## 2. Design

A drift gap of a micromesh chamber defines the number of electron-ion pairs produced in a gas and the time duration of the pulse. To minimize the duration of the pulse the gap can be reduced to zero, but then the pulse amplitude drops to zero due to the negligible ionization in amplification gap. But if the mesh is replaced by a secondary emission solid electrode electrons will be knocked out and develop an avalanche with resulting gain typical to micromegas chambers (~$10^4$). Such detector will have the following advantages:

- very simple construction;
- extremely high radiation resistance;
- short pulse length corresponds to high counting rate and small time resolution.

Fig. 1 shows the fabrication process of such detector. One side copper clad PCB is covered with a photoresistive film 50 –150 μm thick and pillars (200 μm diameter and pitch 2 mm) are obtained by using conventional lithography. The second PCB is placed above. The distance between the plates is fixed by the pillars. The gap is filled with gas at atmospheric pressure and a high voltage is applied about 50 kV/cm. The signal is read out from one of the electrode.

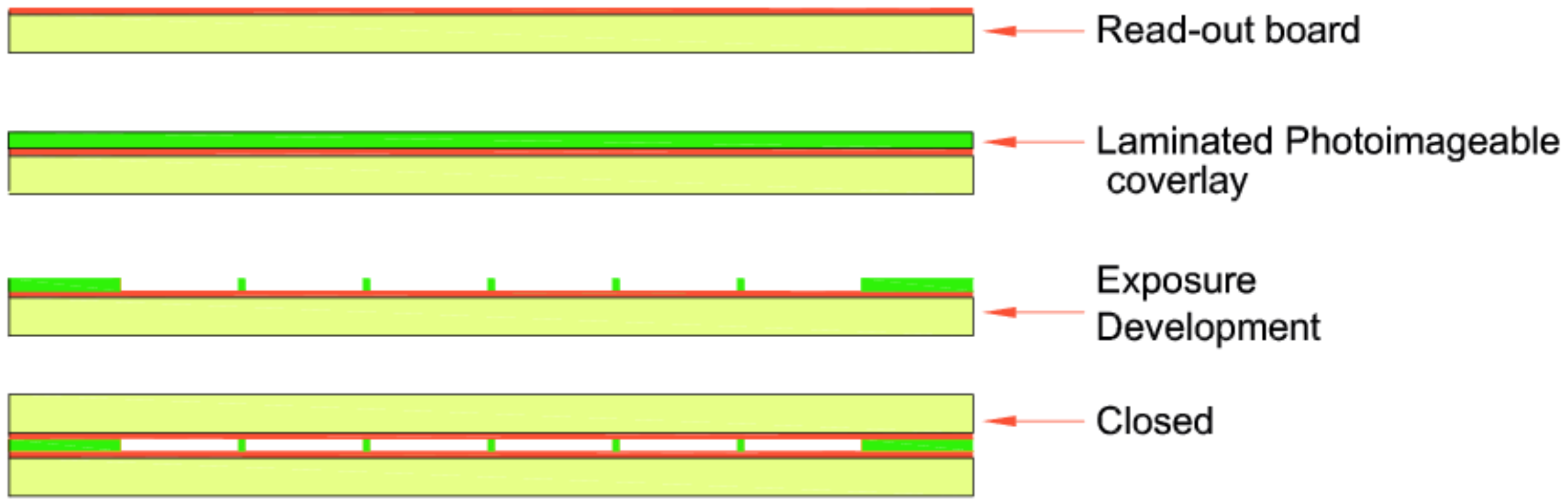

Fig. 1. The fabrication process of the chamber.

The detector operates in the following way. From one electrode a particle knocks out low energy (~ eV) secondary electron which produces in the gas volume avalanche with resulting gain about $10^4$. Probability for the incident particle to ionize gas in the 100 μm gap is very small and besides the ionization will be uniformly distributed along the gap that makes the signal due to ionization small in comparison with the signal of the secondary emission electron.

## 3. Detector test

The studied structure of the detector consisted of 30 mm diameter anode with strips 0.8 mm wide and 1 mm pitch. There were produced 200 μm diameter pillars by the method described above, 150 μm high and 2 mm pitch. The anode was grounded through 1 kΩ. Above the anode a copper clad PCB cathode was placed connected to negative high voltage supplied through 20 MΩ. The electrodes were fixed by supports in a gas volume (fig. 2) through which flows a mixture of Ar+$CO_2$ at atmospheric pressure. The output of the anode was fed to a preamplifier, then to an amplifier and then to a multi-channel analyzer. In all measurements the statistical error in a maximum distribution was not more than 1%.

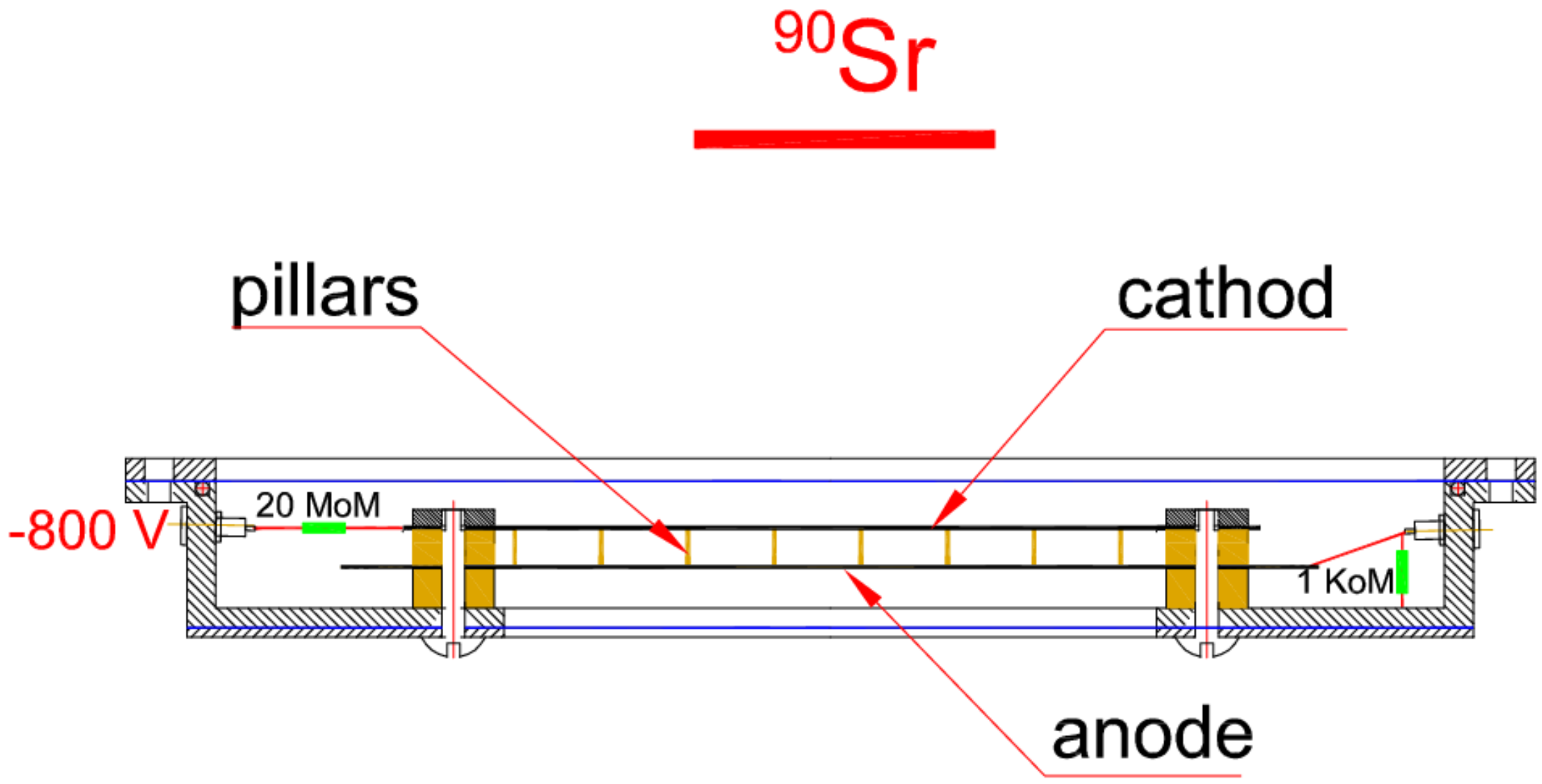

Fig. 2. Set-up of the gas vessel for the measurements.

Obviously the design of the pillars in this chamber can be much simpler than in a micromesh chamber because in this case the cathode is a rigid plate and can be made even of metal. To test it we took a double sided adhesive tape and glued one side to mylar so that the total thickness was 100 μm. Then the pieces were glued to an electrode (fig. 3b), covered by another electrode and then placed into a gas volume (fig. 2). In air such structure was braked at higher voltage then the electrode with pillars (fig. 3a). That can allow to rise high voltage to increase the gain.

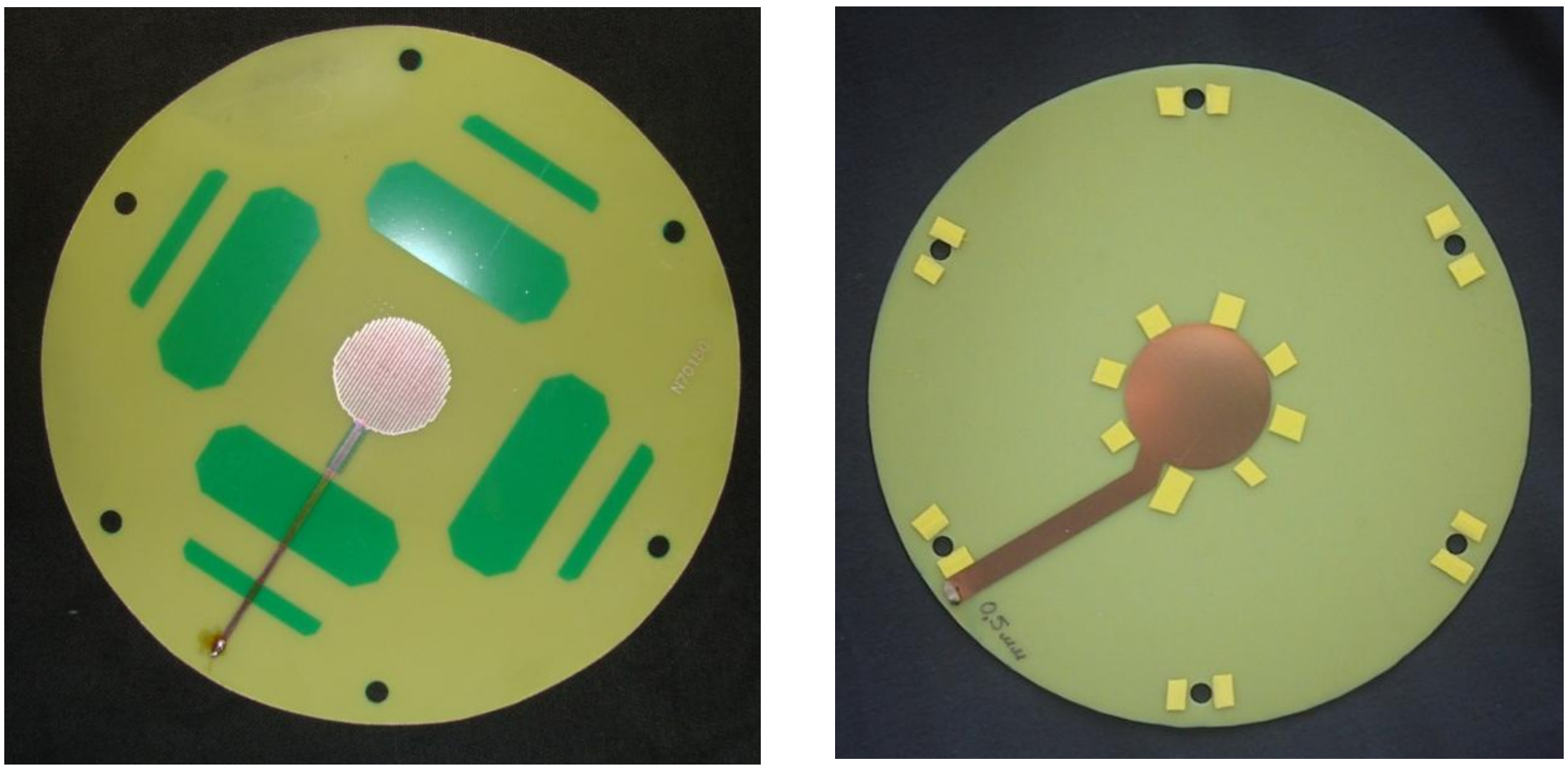

a) b)

Fig. 3. The chamber electrode with a) pillars lithographically produced and b) 100 μm glued spacers (yellow pieces).

## 4. Results

The pulse height amplitude from the secondary emission gas chamber (SEGC) is smaller in comparison with micromesh chamber approximately in the number of electron-ion pairs produced in a drift gap. For gas mixture $Ar+CO_2$ and 3 mm gap width this value about 30. So if for micromesh chamber the pulse height is 1 mV on 50 Om load for SEGC it will be about 30 μV. Fig. 4 shows SEGC pulse after an amplifier (can not exclude some contribution of the amplifier to the pulse length) . The pulse front corresponds to the time of electron collection and the tail is determined by the time collection of ions and practically depends linearly on the gap width.

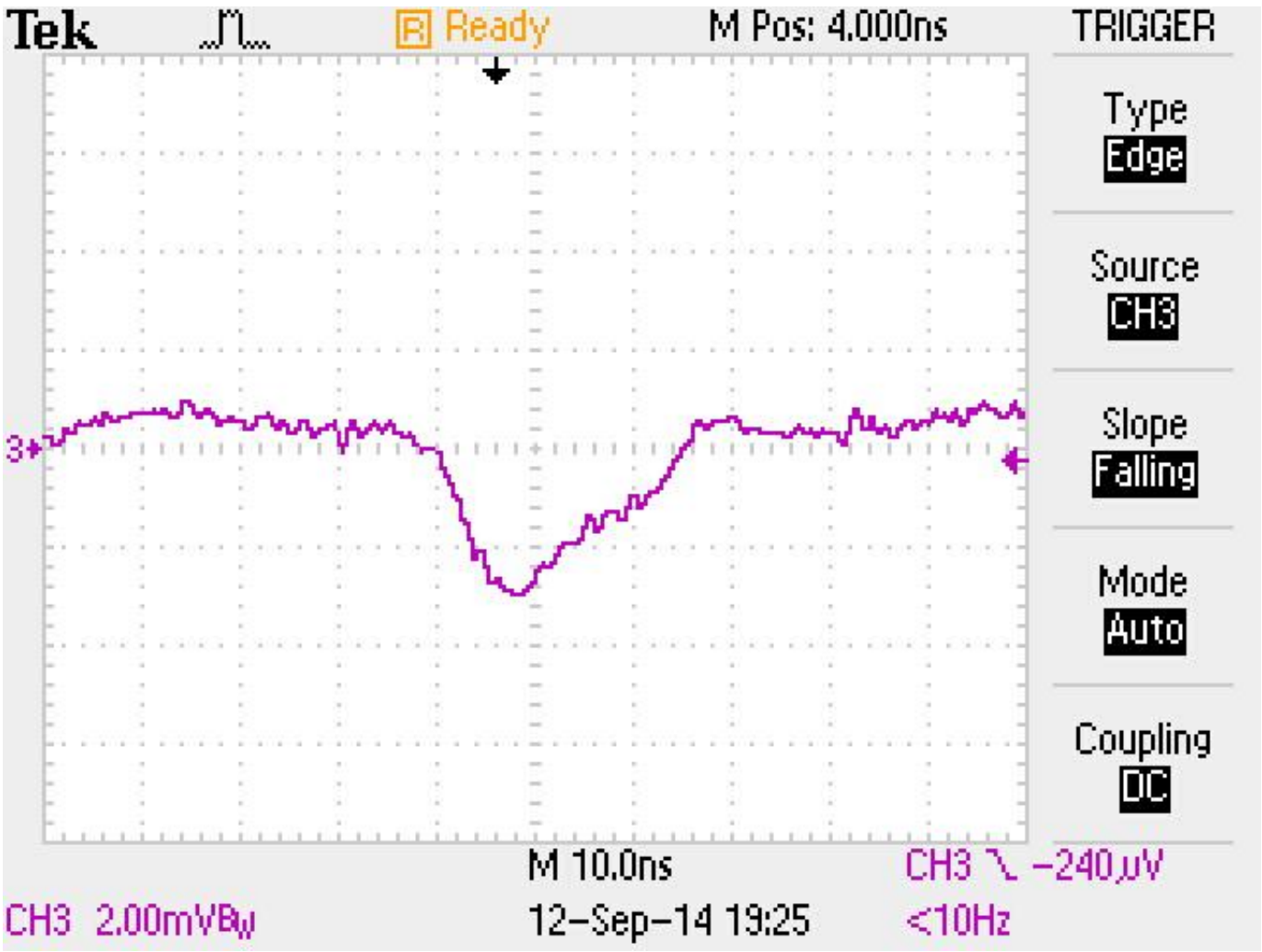

Fig. 4. The pulse shape from $^{90}Sr$ source.

To estimate the gas gain a calibrated charge was fed to the preamplifier. The measurements gave the value about $10^5$ for 150 μm gap, gas mixture Ar+ 7% $CO_2$ and 800 V. The uncertainty is connected to unknown coefficient of the secondary electron emission.

At fixed cathode voltage (-800 V) there was measured a pulse height distribution from the chamber irradiated by $^{90}Sr$ with copper and aluminum cathodes. The results are presented in fig. 5. As one can see the difference as expected is small. A number of materials has appreciably higher coefficient of the secondary emission but as a rule they are dielectrics and can not be considered for high counting rate detectors due to unavoidable polarization.

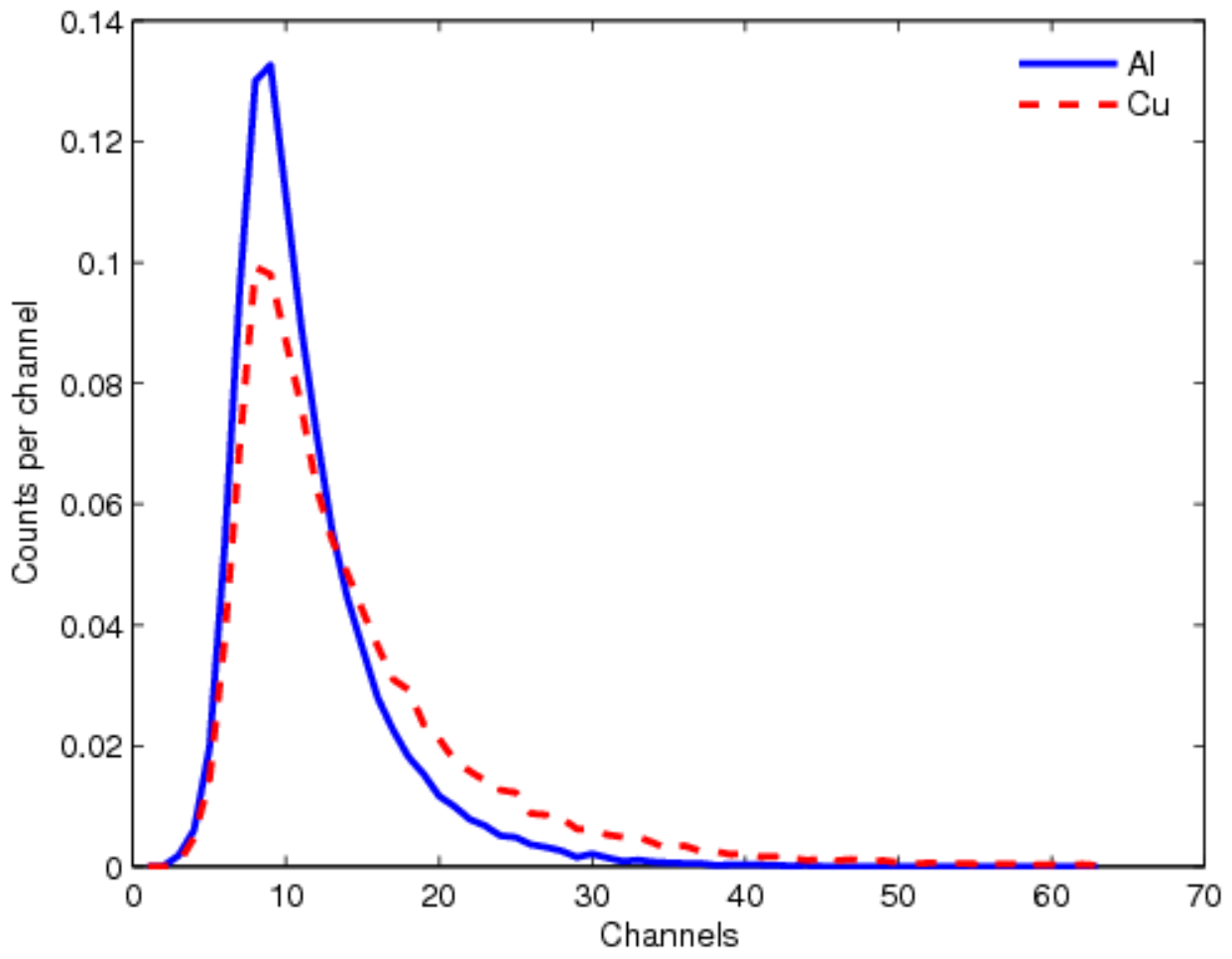

Fig. 5. Normalized pulse height distribution from $^{90}$Sr source for copper (red dotted line) and aluminum (blue solid line) cathodes.

The coefficient of the secondary electron emission can depend on direction of the particle trajectory. To test the isotropy of the chamber response it was irradiated from below. Fig. 6 shows the normalized pulse height distribution obtained for two position of the radioactive source for copper cathode. The absence of any dependence is important for use of such detector in calorimetry.

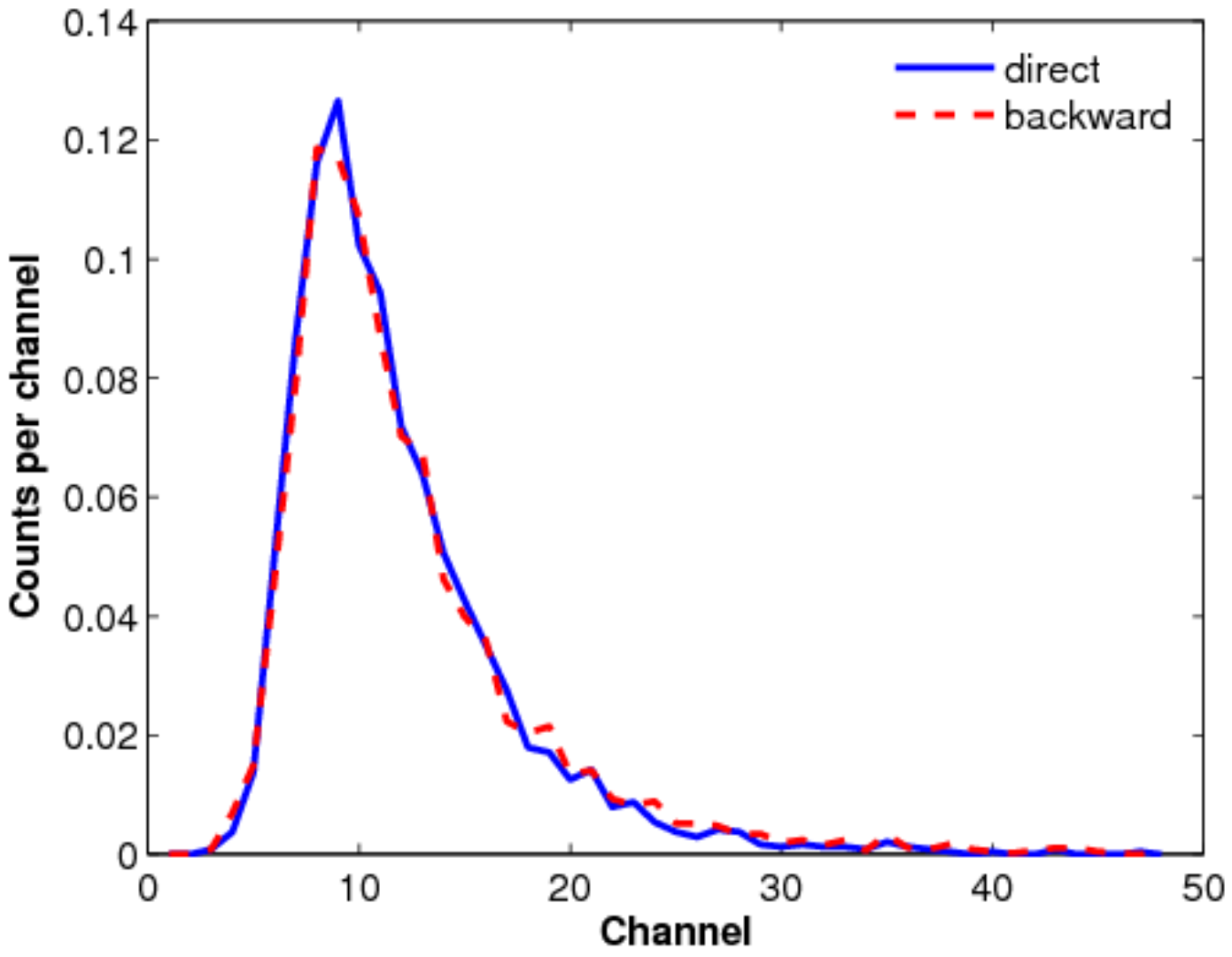

Fig. 6. Normalized pulse height distribution from $^{90}$Sr source placed above the test volume (fig. 2) (red dotted line) and below it (blue solid line).

The dependence of gain on gas mixture in such detector can be different because the coefficient of secondary electron emission can also depend on gas. We measured amplitude distributions for 3 mixtures of Ar+$CO_2$ (7%, 20% and 30% $CO_2$) at high volt-age close to breakdown for each mixture. Fig. 7 shows the typical pulse height distribution that is identical for all mixtures.

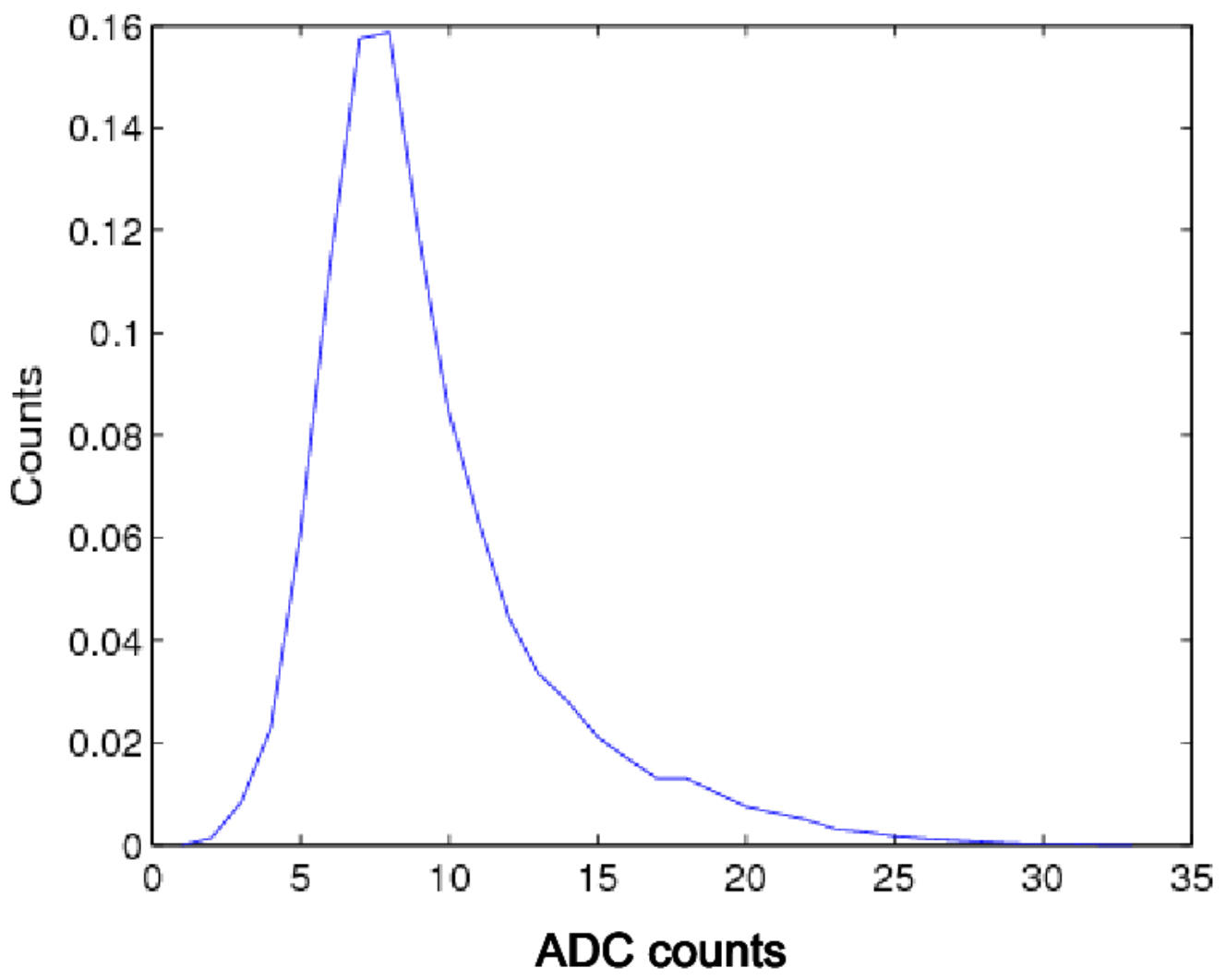

Fig. 7. Normalized pulse height distribution from $^{90}$Sr source for three gas mixtures Ar+C0$_2$ (7 % $CO_2$, 20 % $CO_2$, 30 % $CO_2$). The high voltage was always set close to breakdown.

It is known that the coefficient of the secondary emission depends on many factors such as a particle energy, a particle specie, direction of particle trajectory relative to plane of electrode, electrode material et cetera. The main goal of this work is the development of a calorimeter active elements. Taking into account the complex composition of a nuclear avalanche and intricate nature of the secondary emission the practical method to develop the active elements for calorimetry is to make and test a prototype based on SEGC.

**Conclusion**

The proposed design of the detector has a number of advantages that make it attractive for experiments with very high luminosity where counting rate, radiation resistance and cost are important. The present proof of the principle shows reach potential of the device but the final conclusion can be made after a test of a calorimeter prototype with active elements based on SEGC.

**Acknowledgements**

In conclusion the authors are deeply grateful to G. Britvich, A. Kalinin and M. Soldatov for the help.

## References

1. Serge Duarte Pinto.   rXiv:1011.5529v1.
2. Maxim Titov. arXiv:1008.3736v2.
3. Serge Duarte Pinto. rXiv:1011.5529v1.
4. C. Adloff ,  J. Blaha,  M. Chefdeville et al.  Nuclear Instruments and Methods in Physics Research A 729 (2013) 90.
5. Catherine Adloff, Jan Blaha, Ambroise Espargili`ere et al. arXiv:0901.4927v1.
6. M.C Fouz1. arXiv:1202.5567.
7. C. Adloff, D. Attle, J. Blaha et al. JINST 4 (2009) P11023 , arXiv:0909.3197.
8. A.A. Derevshchikov, V.Yu. Khodyrev, V.I. Kryshkin, et al. Preprint IFVE-90-99, Serpukhov, 1999.
9. G .S . Bitsadze , M .I . Chernetsov, Yu .V. Khrenov et al.  Nuclear Instruments and Methods in Physics Research A 334 (1993) 399.
10. A. Albayrak-Yetkin, B. Bilki, J. Corso et al. arXiv: 1307.8051.